# EXPLORING THE OUTER SOLAR SYSTEM WITH SOLAR SAILING SMALLSATS ON FAST-TRANSIT TRAJECTORIES AND IN-FLIGHT ASSEMBLY OF SCIENCE PAYLOADS

National Academy of Sciences
The Planetary Science and Astrobiology Decadal Survey 2023-2032

## A WHITE PAPER

Slava G. Turyshev[1], Henry Helvajian[2], Louis D. Friedman[3],
Tom Heinsheimer[2], Darren Garber[4], Artur Davoyan[5] and Viktor T. Toth[6]

[1]*Jet Propulsion Laboratory, California Institute of Technology,
4800 Oak Grove Drive, Pasadena, CA 91109-0899, USA*

[2]*The Aerospace Corporation, El Segundo, CA*

[3]*The Planetary Society, Pasadena, CA 91101 USA*

[4]*NXTRAC Inc., Redondo Beach, CA 90277*

[5]*Department of Mechanical and Aerospace Engineering, University of California, Los Angeles*

[6]*Ottawa, Ontario K1N 9H5, Canada*

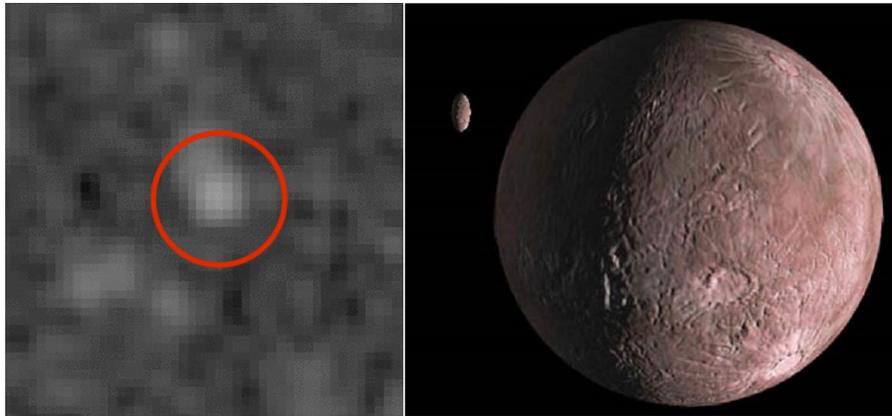

Science impact: Left: Image of Quaoar by New Horizons (2016). Right: To be taken at same distance in ~2030, image of Quaoar and its moon Weywot.

## 1. Faster, Farther and Frequent Opportunities to Explore the Solar System

In the last decade, two new interplanetary technologies have advanced to the point where they enable exciting, affordable missions that reach further and faster into the outer regions of our solar system: (i) Interplanetary smallsats, developed and flown by JPL as MarCO[1] on the Mars InSight mission; and (ii) Solar sails, which utilize solar radiation pressure for propulsion. The successful JAXA-built spacecraft IKAROS[2] demonstrated solar sail technology on a Venus-bound mission launched in 2010. In 2019, the successful orbital demonstration by the LightSail-2[3] flight led by The Planetary Society raised confidence in solar sails and paved the way for two of NASA interplanetary missions, NEA-Scout[4] and Solar Cruiser[5], planned for the near future. Japan is developing OKEANOS[6] as a follow-up to IKAROS for outer planet missions.

The proliferation of smallsats (<50kg) in Earth orbit is testament to their technological maturity. Owing to unprecedented progress with miniaturization of electronics and instrumentation in the last decade, these systems are delivering cutting-edge science and broad commercial applications. Smallsat systems can now deliver impressive capabilities on the distance scales of the solar system. With solar sailing approaching 6-10AU/year velocities, travel times within the solar system and beyond are dramatically reduced (Turyshev et al., 2020a). Importantly, unlike flagship missions, solar sail-driven smallsats are low cost and accessible to many. Smallsats on fast solar system trajectories open new opportunities for affordable solar system exploration, enabling major progress in planetary sciences. Unlike flagship missions, of which only a few can be flown each decade due to resource constraints, smallsats permit many affordable missions to be concurrently designed, built, and flown. Mission selection authorities do not need to painfully cherry-pick a select few from many valid mission proposals. They can instead approve to many. On-going steps to standardize smallsat configurations, interfaces and CONOPS make the use of smallsats for complex missions ever more attractive from the standpoints of cost, risk, schedule and functionality.

In our ongoing NIAC Phase III effort (Turyshev et al., 2020a,b), we perform a system design for a technology demonstration mission (TDM) to fly close to the Sun to achieve high solar system escape velocity, while maintaining sail control and precision navigation. Subsequently, a smallsat will be launched on a rideshare to high Earth orbit. From there, the solar sail could gain energy to enter heliocentric space and begin to spiral in toward the Sun. Spiraling in for about a year to reach a perihelion <0.3 AU would test the sail control and sail materials for a flyby near and around the Sun. After that flyby the spacecraft would achieve a high escape velocity from the solar system, e.g., 6-7 AU/year. This would make it the fastest spacecraft to leave the solar system. The TDM spacecraft would not have RTGs, but would test the low power design, as well as communications, navigation and control. The TDM opens a new approach to exploration of the outer solar system. However, in order for these concepts to mature, several enhancements are required. These are the focus of our ongoing efforts (Turyshev et al., 2020a,b; Aerospace, 2020).

Traditionally, smallsat size and mass limitations are viewed as major constraints. It is expected that they can only carry a limited number of miniaturized science instruments (i.e., magnetometers, cameras, particle detectors, etc.) Limitations in the size of batteries and solar panels is another consequence, which limits on-board power. This leads to reduced on-board computer processing capacity and limited communication capabilities. For Earth-orbiting missions and those in cis-lunar space, where at least solar power is high, some of these limitations can be mitigated (https://aerospace.org/small-satellites). However, for planetary exploration and especially in the outer solar system, the satellite size constrains the type and duration of missions that can be explored. Thus, aperture size plays a big role in the physical design of some instruments (e.g. telescopes, antennas). As a result, the expectation is that interplanetary smallsats find very limited use in deep space to conduct scientific discovery beyond just snapping pictures. This need not be so.

---

[1] https://www.jpl.nasa.gov/cubesat/missions/marco.php
[2] https://global.jaxa.jp/countdown/f17/overview/ikaros_e.html
[3] https://www.planetary.org/explore/projects/lightsail-solar-sailing/
[4] https://www.nasa.gov/content/nea-scout/
[5] https://en.wikipedia.org/wiki/Solar_Cruiser
[6] https://en.wikipedia.org/wiki/OKEANOS

Before deep space smallsat missions can fly, three technical enhancements are needed. These form the basis for the present proposal: 1) proof of the ability of solar sails to propel smallsats to 6-10 AU/year by perihelion passage; 2) proof of the ability of smallsats to operate throughout the mission environment – spiral down to perihelion, perihelion passage, flyout to distant targets; and, 3) enhancement of smallsat mission capability by deep-space smallsat swarms that share mission functions by information transfer and physical rendezvous and docking.

To address these challenges, we want to 1) maintain fast-transit capability, with solar sails offering the only solution that is cost-affordable and possible in the near-term; and 2) provide planetary researchers with more exhaustive scientific data. Given that fast transit via sailcraft propulsion becomes impractical for mass approaching 50 kg, we explore the use of inflight robotic assembly of modular segments delivered on multiple fast moving sailcraft accelerated by solar radiation pressure. Other supporting arguments are: (i) Solar sailing consumes no propellant to reach high velocities but forces spacecraft designs to be low mass, thus significantly lowering the spacecraft cost (~$10m.). (ii) The solar sail – smallsat architecture to be produced in this study could be ready for flight implementation in less than a decade (Turyshev et al., 2020a,b).

## 2. Potential Science Impact

Our concept has the potential to enable wholly new classes of missions, offering significant advantages by providing great leaps in capabilities for NASA and the greater aerospace community. Turyshev et al. (2020c) discuss some of the exciting science objectives for fast-moving spacecraft that could lead to transformational advancements in the space sciences in the coming decade, including the following investigations:

- *Studying Quaoar, other Kuiper Belt Objects (KBO):* The Kuiper belt is a disc-shaped region beyond the orbit of Neptune, extending to 50 AU from the Sun. The recently discovered dwarf planets – Haimea, Makemake, Eris, and Quaoar – all provide interesting targets for exploration. These objects orbit the Sun at the very edges of the solar system at distances ranging 40–90 AU. A mission reaching out to the outer solar system presents a unique opportunity to fly by a large KBO. Quaoar is one of the most interesting KBOs. It is in a transition between the large, volatile-dominated, atmosphere-bearing planetesimal and the typical mid-sized volatile-poor object. For most of its history, it had a methane atmosphere, but it is now in the last stages of losing it. Most likely, its surface is patchy in methane frost with the methane being mostly cold trapped near the poles or in craters. The processes related to atmospheric loss in the outer solar system are poorly known, so Quaoar offers an interesting opportunity to see the process in its late stages. Given its size, Quaoar may have ancient cryovolcanic flows on the surface, offering clues on its history. To investigate these processes, we can make full global imaging in broadband colors, which may be achieved with a swarm of sailcraft. We may study Quaoar's interior using one of the sailcraft as an impactor. Imaging the crater would be an interesting probe into surface conditions. Plume spectroscopy could explore subsurface composition. Sailcraft may be used to carry out these investigations and that of many other KBOs.

- *Exploring Enceladus*. Enceladus, Saturn's potentially life-supporting moon, is an exciting scientific target that has one of the most compelling and challenging environments. One of the top priorities of a future mission to Enceladus will be the investigation of the composition and chemistry of its subsurface ocean.  Such a mission must determine the thickness and dynamics of the Enceladus' ice shell, characterize the surface geology. An imaging system would be an important part of the mission's instrument suite. On a "flyby" mission, a "staring" camera is used when the object is a long range away and a "push broom" camera is used when the spacecraft is near the target. Such a camera can also be used for spacecraft navigation by imaging solar system objects against background stars. Such an autonomous on-board navigational capability minimizes the need for DSN tracking, reducing the cost of mission operations and leading to totally autonomous navigation for planetary missions.  The resulting accuracy improvement will lead to advancements in many geophysical investigations. This approach recognizes the driving requirements for outer solar system imaging (low light, fast fly-bys) and relies on new generations of sensors capable of addressing those conditions.
- *Rendezvous with Interstellar Objects (ISO)*, such as 1I/'Oumuamua and 2I/Borisov that were detected in 2017 and 2019, respectively. It is likely that many such objects pass undetected through the solar system every year. Photographing or visiting these objects and conducting *in situ* exploration would allow us to learn about the conditions in other planetary systems without sending interstellar probes. Sailcraft on high-energy trajectories provide a unique opportunity to directly study ISOs transiting through the solar system. The scientific return from such investigations is invaluable, as comparative studies between an ISO sample return with solar system asteroid and comet sample returns can help us understand the conditions and processes of solar system formation and the nature of the interstellar matter—the first priority question listed in the Planetary Sciences Decadal Survey[7].  With many new ISOs expected, this topic should be of the highest priority for the new decade. Small sun-propelled sailcraft are the only means of exploration we have that could catch up with ISOs moving at tens of km/s and be inexpensive enough to be on standby in Earth orbit. This capability is within reach today and should be considered for planetary exploration.
- *Reaching the Solar Gravitational Lens (SGL) for Imaging of Exoplanets*. Our ultimate objective is the imaging of extrasolar terrestrial planets combined with spectroscopy, which probably is the single greatest remote sensing result that we can contemplate in terms of galvanizing public interest and support for deep-space exploration. The SGL offers the means to produce high-resolution, multipixel images of exoplanets. A meter-class optical telescope with a modest coronagraph operating in SGL's focal region, beyond 650 AU, can yield a $200\times200$-pixel image of an Earth-like exoplanet at 30 pc in just 12 months, not possible otherwise. Turyshev et al. (2020a,b), have developed a new mission concept to deliver optical telescopes to the SGL's focal region and then to fly along the focal line to produce high resolution, multispectral images of a potentially habitable exoplanet.  The multi-smallsat architecture uses solar sailing and is designed to perform observations of multiple planets in a target extrasolar planetary system. It allows to reduce integration time, to account for target's temporal variability, which helps to "remove the cloud cover".  If this mission can provide spectroscopic proof of life on an exoplanet, it would qualify as one of the greatest scientific discoveries in human history.

This concept has the potential to generate enthusiasm with the planetary science community as it provides unique opportunities for frequent and inexpensive access to the targets of interest in the deep solar system. As we are heavily relying on the COTS components developed in the industry, it may lead to new business models where industry (i.e., a GPS constellation and communication links from Mars, etc.), helps to build advocacy to support it within NASA and in the greater aerospace community. This interest may lead to public-private partnerships to explore the vast region of the outer solar system with capable smallsats on fast solar system transit trajectories.

---

[7] https://www.nap.edu/catalog/13117/vision-and-voyages-for-planetary-science-in-the-decade-2013-2022

### 3. In-flight Assembly of Capable Science Payloads

In-flight (as opposed to on-orbit around Earth or in cis-lunar space) autonomous assembly is plausible. We can rely on in-flight autonomous assembly where sizable instruments (for instance, a meter-class optical telescope – an optical telescope being one of the most challenging instruments to assemble, as most other instruments do not rely on submicron tolerances for optical accuracy and thus are much more easily assembled without direct supervision) may be constructed from modular parts delivered by a group of smallsats (< 20 kg) which are placed on fast solar system transfer trajectories while being accelerated by solar sail propulsion to velocities of ~10 AU/yr.

This concept provides the planetary science community with inexpensive, frequent access to distant regions of the solar system with flexible, reconfigurable instruments and systems that are assembled in flight. It permits faster revisit times, rapid replenishment and technology insertions, longer mission capability with lower costs; higher resolution maps and global monitoring of the target surface. It also increases capabilities while avoiding the impact of a catastrophic failure of a flagship type mission, via the use of modular, redundant architectures and the proliferation of sensing instrumentation throughout the solar system.

Demonstration of the capabilities of small and inexpensive sailcraft, placed on fast solar system transit trajectories and reaching speeds of ~10 AU/yr (Turyshev et al., 2020a,b) is one of the primary objectives of the ongoing NIAC Phase III effort (Aerospace, 2020; Xplore, 2020). A cruise velocity of 10 AU/yr enables a sailcraft to reach Pluto in 4 years. The trajectory may start from Earth orbit, spiraling in toward the Sun and then, after passing through perihelion, transition to a high-energy hyperbolic trajectory to the outer solar system and beyond. But this smallsat is minimally capable. Consequently, for some missions it needs to be joined with other fast travelling functional units to form the mission spacecraft. A key element in this investigation is the space architecture and CONOPS of how to parse out the functional elements of a scientific discovery spacecraft into elements that can navigate to the location where in-space assembly transpires.

The architecture employs multiple smallsats and ultrafast solar-propelled trajectories that create a series of increasingly ambitious missions even out of the ecliptic. This capability can be used to explore targets outside the ecliptic plane, thereby enabling a large set of interesting science objectives (Turyshev et al, 2020a,c). As yet unimagined missions become possible, once the science community recognizes the ease with which this approach can reach "anywhere and everywhere".

The mission architecture relies on 5 primary phases:

<u>Phase 1</u>: *Affordable access to space*: Using rideshare opportunities we launch multiple 6U smallsats. Each smallsat is equipped with a solar sail, a small solar array and small comm antenna to support a radio link up to 2 AU. In addition, each smallsat carries segments of the key scientific instruments that will be assembled in space.

<u>Phase 2</u>: *Solar flyby, trajectory injection*. The fleet of smallsats follow the same trajectory toward solar perihelion. Position/velocity differences are trimmed with the sails – some sailcraft fractionally slow down, some shift in (*x,y*) direction perpendicular to the inbound trajectory. Eventually, all the sailcraft are aligned in sequential fashion with (*x, y, $v_x$, $v_y$*) errors minimized. After perihelion, further trajectory trimming transforms the cluster into a close-knit interoperable mission unit.

Phase 3: *Sail jettisoning and docking*. Outbound, solar sails are used to trim velocity with the formation going to "tight formation". Once the sails are jettisoned the "wingless" modules operate on batteries and ion propulsion. As needed, some of the modules initiate proximity operations and docking. The mooring operations are assisted by magnets; positioning/instabilities are compensated via reaction wheels with desaturation via thrusters. The fleet then consists of some physically connected modules, and some that are flying in formation to provide needed mission capability.

Phase 4: *Forming a telescope*. While other sensing instruments can be formed using the architecture described above, in this proposal we take on the most challenging instrument, an optical telescope. Once the rigid cluster is formed, a reconfiguration forms a meter-class telescope. In the baseline design, one of the modules would carry an RTG, while others carry batteries so upon cluster formation, the RTG begins to recharge the batteries for the integrated system. The telescope segments are calibrated and adjusted in relation to each other to submicron tolerances. The topology change produces a boom that either holds the imaging camera or a deformable mirror for minute wavefront corrections to create a diffraction-limited telescope.

Phase 5: *Telescope operations*. A meter-class telescope has two purposes. It is used for 1) imaging and observations, as well as for 2) optical COMM downlink. The downlink and imaging positioning are handled by the cooperative action of the on-board reaction wheels and ion propulsion. In this configuration, all the resources (power, computing, propulsion, etc.) are shared and optimized.

The concept allows fast access to the distant region of the solar system by delivering spacecraft that could have an exhaustive list of instrumentation to a target designation. It also permits the release of instrumentation during the flyby (such as a compact impactor to allow for spectroscopic investigations of a target body, such an interstellar object). This is made possible by the ability to conduct instrument assembly from a set of modular components, all individually delivered by sailcraft accelerated to large velocity (~10AU/yr) and then physically aggregated to form a larger and more capable spacecraft which continues to fly at ~10AU/yr. These larger space systems could afford medium-size radio antennae or a large optical telescope aperture.

This concept reduces cost, allows repurposing, and increases the number of possible missions. It enables new missions by innovative aperture assemblies in space without need for large flagship-class satellites and costly launchers. For planetary science, in-flight assembly allows for the construction of large apertures, leading to higher resolution imaging of faint objects and in increased mission duration during a fast fly-by trajectory. The New Horizons mission to Pluto took ~10 years to reach its destination for a 16-hour mission. With the approach discussed here, we can repeat the same feat in less than half the time for a much longer science mission.

Our approach addresses a clearly identified gap for an affordable COMM and imaging data from deep regions in the solar system. Planetary science is a clear winner and is poised to benefit from larger (~1m) telescope apertures, antennas, magnetometers, particle detectors, etc., delivered to deep space for investigations on fast flyby trajectories. Other benefits include in-flight assembly of telescope via deployment of subassemblies of optical quality instruments by ride-share opportunities; precision in-orbit assembly; wavefront control optics; and applications of AI/machine learning tools to autonomously assemble mission-ready optical instruments in space.

## 4. Credibility and Overall Technology Readiness

As an example, we consider an autonomous assembly of a modest, meter-class planetary camera in space. Such systems could enable many remote investigations of faint objects conducted from a spacecraft on fast flyby trajectories. These systems are currently limited to apertures of ~30cm.

With our approach their apertures may be increased up to 3 times that size, thus increasing their sensitivity at least by an order of magnitude.

In recent years there has been a desire to develop space-based optical telescopes with very large primary apertures (over 20 m in diameter). Currently the largest aperture under development is that of the JWST (James Webb Space Telescope) with a diameter of 6.5 m. JWST represents a major shift in telescope design due to its use of a deployable primary mirror.

Designs for even larger apertures exist. To overcome the limit of currently available launch vehicle fairings, there is a shift away from monolithic mirrors. One method would be to form the telescope in orbit by means of the autonomous assembly of small independent spacecraft, each with their own mirror segment. This way, a telescope with a large, segmented primary mirror can be constructed. Furthermore, if each of these mirrors is manufactured to have an identical initial shape, which can then be adjusted upon assembly, a substantial reduction in manufacturing costs can be realized while also increasing mission redundancy. Such an approach may allow for the practical and cost-effective realization of future large aperture (~100 m diameter) space telescopes and may have broader application in other space missions that require on-orbit assembly.

To this end, AAReST[8] (Autonomous Assembly of a Reconfigurable Space Telescope) is being developed to deploy a multisegment astronomical telescope. When flown in 2021, it will demonstrate autonomous assembly and reconfiguration of a space telescope based on multiple mirror elements, including the use of adaptive mirrors. It will also demonstrate the capability of providing high-quality (astronomical) images. Additionally, the DeMi[9] (Deformable Mirror Demonstration Mission, to be flown in 2020) will demonstrate optical phase control with deformable mirror. An additional related project is the NASA-funded iSAT study[10] for large (5-15m+) aperture assemblies. These projects lend foundational technologies and credibility to our approach.

Prior studies focused on the needs of the astrophysics community for extremely large apertures. We take the results of these studies, add solar sailing technology for fast transit through the solar system, reduce the mass of the individual spacecraft and scale down to aperture requirements to address the needs of planetary science investigations where access to modest size meter-class apertures, is critical, but not currently feasible. We focus on connecting the segments and the in-flight assembly of these instruments to deliver workable scientific instruments to the far reaches of the solar system. We focus on imaging applications and optical communication downlinks.

Compared to the earlier work, our focus is more wide-ranging: to perform assembly by relying on precision autonomous navigation, proximity operations and docking to build spacecraft and payloads based on cooperative principles (see, e.g., https://www.satelliteconfers.org/). This approach speeds up technology refresh and enables repairs if needed: i) AI-managed rendezvous, docking/assembly/reconfiguration with minimal ground support, ii) Use of robotics to assemble aperture, electro-permanent magnetics to hold in place, iii) Shared-launch, economies of scale, modular design that accelerates the technology refresh time and reduces total ownership cost.

Our approach relies on the following facts: 1) Modern smallsats are capable of precision navigation, proximity operations and docking. 2) A 1-m aperture, 0.6-1.0-micron, optical telescope assembled from ~30 cm segments technologically is more plausible that assembling a 30 m telescope with $10^3$ segments. 3) Significant advances have been made using origami folding structures. 4) Deformable structures with optical wavefront control and analysis algorithms exist. 5) There have been large strides in the field of robotics and miniaturized actuators coupled with distributed control algorithms. 6) There is a large commercial-of-the-shelf (COTS) market of nanosatellite subsystems technologies with ~10-year reliability. Thus, we feel confident that the new feature, which is in-flight assembly of functioning segments while moving fast on sailcraft platforms is possible. These segments are based on a novel concept utilizing "LEGO-block design" (see, https://www.novawurks.com/) and "in flight assembly" with origami structures, most of which already exist piecewise with some being demonstrated in Earth orbit.

---

[8] https://directory.eoportal.org/web/eoportal/satellite-missions/a/aarest  
[9] https://directory.eoportal.org/web/eoportal/satellite-missions/d/demi  
[10] https://exoplanets.nasa.gov/exep/technology/in-space-assembly/iSAT_study/

All technologies exist separately and have reasonable readiness levels. We put the segments of the instruments on the fast-moving sailcraft and perform the assembly of a larger and more capable spacecraft in flight. This aggregation approach achieves new capabilities from small segments.

## 5. Conclusions

We discuss an approach that enables a new generation of science missions that could be undertaken by NASA in collaboration with its international and industry partners. The concept of clustered smallsats that form larger and more capable systems could be used to establish new capabilities to explore deep regions of the solar system. As a first step we have proposed a pathfinder TDM, now being studied as a potential private-public partnership in the ongoing NIAC Phase III (Turyshev et al., 2020b). This mission would launch on a rideshare to Earth escape velocity, spiral towards the Sun and then achieve a hyperbolic trajectory at ~6-8 AU/year – twice as fast as Voyager 1. The purpose would be to demonstrate the fast-solar system exit smallsat/sail capabilities.

After that, the proven fast transit capabilities could be combined with in-flight autonomous assembly to demonstrate rapid transit "probes" to a solar system target, ultimately extending out to the KBOs/dwarf planets. These would be low-cost flybys with an advanced primary sensor on board. With solar sails and solar perihelion exit velocities of 20 AU/yr, Jupiter is reached in 3 months, Saturn in 6 months, Pluto in less than 2 years, and the outer edge of the Kuiper belt in 3 years.

The presented approach would accelerate and expand scientific exploration of the solar system and beyond. Further NASA investments focused on low cost, mass-producible smallsats, as well as sailcraft technology and inflight spacecraft/payload reconfiguration solutions would bring this concept to rapid fruition. This would leverage ongoing work in the private space industry and many synergistic industrial robotics efforts. This approach would encourage creation of robust public-private partnerships to benefit many planetary science investigations. Such missions will be more frequent, more capable, and less expensive, encouraging the next generation of scientists.

**Acknowledgements.** The work described here in part was carried out at the Jet Propulsion Laboratory, California Institute of Technology, under a contract with the National Aeronautics and Space Administration. ©2020. All rights reserved.